\documentclass[twocolumn,showpacs,preprintnumbers,amsmath,amssymb,floatfix]{revtex4}
\usepackage{graphicx}

\begin{document}


\title{Kondo resonances and anomalous gate dependence of electronic conduction in single-molecule transistors}

\author{L.H. Yu$^{1}$, Z.K. Keane$^{1}$, J.W. Ciszek$^{2}$, L. Cheng$^{2}$, J.M. Tour$^{2}$, T. Baruah$^{4,5}$, M.R. Pederson$^{4}$, D. Natelson$^{1,3}$}

\affiliation{$^{1}$ Department of Physics and Astronomy, Rice University, 6100 Main St., Houston, TX 77005}
\affiliation{$^{2}$ Department of Chemistry and Center for Nanoscale Science and Technology, Rice University, 6100 Main St., Houston, TX 77005}
\affiliation{$^{3}$ Department of Electrical and Computer Engineering, 
Rice University, 6100 Main St., Houston, TX 77005} 
\affiliation{$^{4}$ Center for Computational Materials Science, Code 6390-Naval Research Laboratory, Washington, DC 20375-5435}
\affiliation{$^5$Department of Physics, SUNY, Stony Brook, NY 11794}

\date{\today}

\begin{abstract}
We report Kondo resonances in the conduction of single-molecule
transistors based on transition metal coordination complexes.  We find
Kondo temperatures in excess of 50~K, comparable to those in purely
metallic systems.  The observed gate dependence of the Kondo
temperature is inconsistent with observations in semiconductor quantum
dots and a simple single-dot-level model.  We discuss possible
explanations of this effect, in light of electronic structure calculations.
\end{abstract}

\maketitle

\newpage

In the Kondo Hamiltonian\cite{Kondo64PTP}, one of the most
well-studied many-body problems in physics, an unpaired spin localized
in a singly occupied electronic level is coupled via tunneling to an
electronic bath.  On-site Coulomb repulsion forbids real double
occupancy of the level, but virtual processes favor antiferromagnetic
exchange between the local spin and the electronic bath.  As $T$ is
reduced below a characteristic Kondo temperature, $T_{K}$, these
exchange processes ``screen" the local moment.  The Kondo
problem has undergone a resurgence, with atomic-scale studies of Kondo
physics by scanning tunneling microscopy
(STM)\cite{MadhavanetAl98Science,NagaokaetAl02PRL} and the realization
of tunable Kondo systems in semiconductor quantum
dots\cite{GGetAl98Nature,CronenwettetAl98Science,GGetAl98PRL,vanderWieletAl00Science}.
With the recent development of single-molecule transistors (SMTs)
based on individual small molecules\cite{ParketAl00Nature}, Kondo
systems now include organometallic
compounds\cite{LiangetAl02Nature,ParketAl02Nature} and fullerenes with
normal\cite{YuetAl04NL} and ferromagnetic\cite{PasupathyetAl04Science}
leads.

In this Letter, we report Kondo physics in SMTs incorporating
transition metal complexes designed to contain unpaired electrons.  As
a function of gate voltage, $V_{\rm G}$, we observe transitions from
Coulomb blockade conduction to Kondo conduction, manifested as a
strong peak in the differential conductance, $G \equiv \partial I_{\rm
D}/\partial V_{\rm SD}$, at zero bias, $V_{\rm SD} = 0$ in one charge
state.  At fixed $V_{\rm G}$, the temperature dependence of the
conductance peaks' amplitudes and widths agree well with the expected
forms for spin-1/2 Kondo resonances.  Observed SMT Kondo temperatures
are $>\sim$~50~K, comparable to those in purely metallic Kondo
systems.  We find that $T_{\rm K}(V_{\rm G})$ is strongly inconsistent
with the simple model of Kondo physics as seen in semiconductor
devices\cite{GGetAl98PRL,vanderWieletAl00Science}.  We discuss
explanations for this anomalous gate dependence in light of
spin-resolved electronic structure calculations of the complexes.

\begin{figure}[h!]
\begin{center}
\includegraphics[clip, width=8cm]{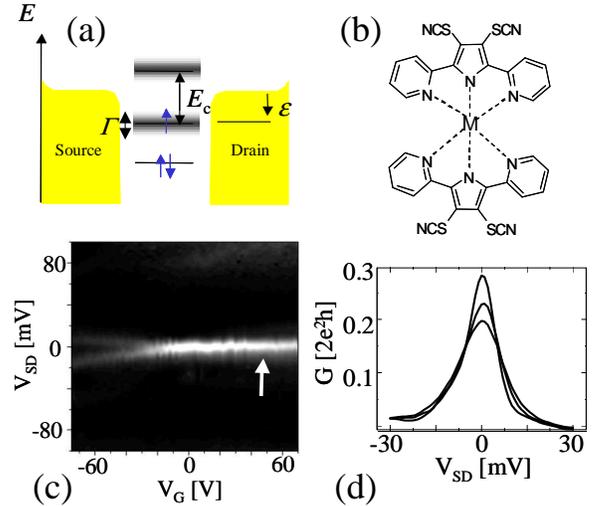}
\end{center}
\vspace{-5mm}
\caption{\small (a) Energy level diagram of a typical SET in the Kondo
regime. (b) Structural formula of transition metal complex before
self-assembly. (c) Map of $G(V_{\rm SD},V_{\rm G})$ showing transition 
to Kondo resonance at 5~K.  Brightness scales from $G=0$ (black) to
$0.3 \times 2e^2/h$.  (d) $G$ vs. $V_{\rm SD}$ 
at $V_{\rm G}=50$~V (as indicated with arrow in (c)) at (top
to bottom) 5~K, 16~K, and 30~K.}
\label{fig1}
\vspace{-5mm}
\end{figure}

Figure~\ref{fig1}a shows the simple single-level picture often used
to describe Kondo physics in single-electron transistors (SETs).  The
intrinsic width of the single particle level is $\Gamma\equiv
\Gamma_{\rm S}+\Gamma_{\rm D}$, determined by overlap of the
single-particle state with the conduction electron states of the
source and drain.  The charging energy $E_{\rm c}$ is the Coulomb
cost of adding an extra electron to the molecule.  The energy
difference between the singly occupied level and the
source/drain chemical potential is $\epsilon$, which is zero at charge
degeneracy and varies linearly with $V_{\rm G}$.

Figure~\ref{fig1}b shows the structure of the neutral transition metal
complexes measured in this study, as-synthesized.  Two planar,
conjugated ligands provide an octahedral coordination (compressed
along the interligand ($z$) axis) of a transition metal ion, M.  We
have examined complexes with M~=~Co, Cu, and Zn, as well as individual
ligands and alkane chains.  The Kondo devices in this Letter contain
Co(II) and Cu(II).  The as-synthesized molecules are characterized by
EPR, SQUID, x-ray diffraction, cyclic voltammetry (CV), IR, and Raman
spectroscopy.  The Co(II) complex is high spin (3/2), with CV in
solution confirming easily accessible Co(II)$\leftrightarrow$Co(III)
redox transitions.  The Cu(II) complex is spin 1/2, with CV supporting
easily accessible Cu(II)$\leftrightarrow$Cu(I) redox transitions.
These complexes self-assemble on Au in tetrahydrofuran (THF) through
loss of the -CN moieties and formation of Au-S covalent
bonds\cite{CiszeketAl04JACS,YuetAl04PRL,supp}.  In principle, the
ligands and remaining -SCN groups can also change their charge state.

Devices are fabricated using Au source and drain electrodes on
degenerately doped $p+$ silicon substrates which are used as the gate,
with 200~nm of gate oxide.  Source/drain electrodes are created by the
controlled electromigration\cite{ParketAl99APL} from lithographically
defined Ti/Au (1~nm/15~nm thick) constrictions that are exposed for 1
minute to oxygen plasma after liftoff, followed by self-assembly of
molecules in solution (2~mM in THF) for 48 hours.  After rinsing and
drying with dry nitrogen, substrates are placed in a variable
temperature probe station for electromigration and measurement, with
fabrication statistics comparable to past
results\cite{YuetAl04NL,YuetAl04PRL}.

DC measurements of $I_{\rm D}-V_{\rm SD}$ are performed with the
source electrode grounded, at various $V_{\rm G}$.  Differential
conductance is computed by numerical differentiation, with spot checks
by lock-in amplifier techniques.  Device stability limits $|V_{\rm
SD}| \le 100$~mV, while gate oxide limits $|V_{\rm G}|\le 100$~V.  We
consider only devices with significant gate response such that a
charge transition is detected in maps of $G(V_{\rm SD},V_{\rm G})$, to
distinguish molecule-based effects from artifacts (e.g. metal
nanoparticles).

Figure~\ref{fig1}c shows a conductance map for a typical device
exhibiting a Kondo resonance.  We have sufficient gate coupling to
observe only a single charge degeneracy point for each device.  From
the slopes of the boundaries of the blockaded regime, one can estimate
the constant of proportionality between changes in $eV_{\rm G}$ and
$\epsilon$ as expected from the model of Fig.~\ref{fig1}a.  For the
device shown, $\delta \epsilon \approx (C_{\rm G}/C_{\rm tot})e \delta
V_{\rm G} \approx~10^{-4}~e\delta V_{\rm G}$.  The average value of
this coefficient is $10^{-3}$.  The width of the charge degeneracy
resonance on the blockaded side of the degeneracy point sets an upper
limit on $\Gamma$ for the level participating in the redox
state change.  The $T \rightarrow 0$ limit of that width is
proportional to $\Gamma$, while the degeneracy resonance is thermally
broadened at finite temperature.  Typical $\Gamma$ values inferred
from 5~K data in these devices are 3-30~meV.  We note that the edges
of Coulomb blockade diamonds, while usually distinct in the non-Kondo
charge state, are often much weaker or apparently absent in the Kondo
charge state, as also mentioned in Ref.~\cite{PasupathyetAl04Science}.

\begin{figure}[h!]
\begin{center}
\includegraphics[clip, width=8cm]{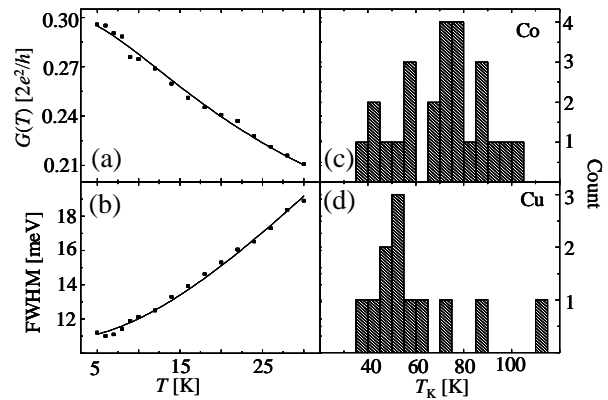}
\end{center}
\vspace{-5mm}
\caption{\small (a) Temperature dependence of Kondo resonant 
peak height for the device of Fig.~\ref{fig1}c at $V_{\rm G}=50$~V.
Solid line is the expected semiempirical functional form for spin-1/2 Kondo,
with $T_{\rm K}=69$~K. (b) Temperature dependence of Kondo peak FWHM for
the same device and $V_{\rm G}$, with fit to expected functional
form for spin-1/2 Kondo.  Setting FWHM in the low-$T$ limit to $2 k_{\rm B}T_{\rm K}/e$ gives $T_{\rm K}=65$~K. 
(c,d) Histograms of $T_{\rm K}$ as inferred from peak widths for Co and Cu complex devices, respectively.
}
\label{fig2}
\vspace{-5mm}
\end{figure}

Fig.~\ref{fig1}d shows $G(V_{\rm SD},V_{\rm G} = {\rm 50~V})$ at three
temperatures.  The resonant peak decreases in magnitude while
increasing in width as $T$ is increased.  Figs.~\ref{fig2}a,b show the
peak height and width, respectively, as a function of temperature.
The solid line in (a) is a fit to the semiempirical
expression\cite{GGetAl98Nature} for the spin-1/2 Kondo resonance in
conduction, $G(T)=G_{\rm c}(1 + 2^{1/s - 1}\frac{T^{2}}{T_{\rm
K}^{2}})^{-s}$, with $s = 0.22$.  After the subtraction of a smooth
background conductance, the adjustable parameters are the overall
conductance scale, $G_{\rm c}$, and $T_{\rm K}$.  Similarly, the solid
line in (b) is a fit to the expected form\cite{NagaokaetAl02PRL} for
the full-width at half-maximum, ${\rm FWHM} = \frac{2}{e} \sqrt{(\pi
k_{\rm B}T)^{2}+ 2(k_{\rm B}T_{\rm K})^{2}}$, where the only
adjustable parameter is $T_{\rm K}$.  The numerical coefficients are
accurate to within a factor of order unity; other
observations\cite{vanderWieletAl00Science,ParketAl02Nature} find
better quantitative consistency with $T_{\rm K}$ as inferred from
$G(V_{\rm SD}=0,T)$ if the FWHM is assumed to be $\approx (2 k_{\rm
B}T_{\rm K})/e$.  We adopt this latter assumption, and find good
quantitative consistency between $T_{\rm K}$ values inferred from
resonance height and width.

\begin{figure}[h!]
\begin{center}
\includegraphics[clip, width=8cm]{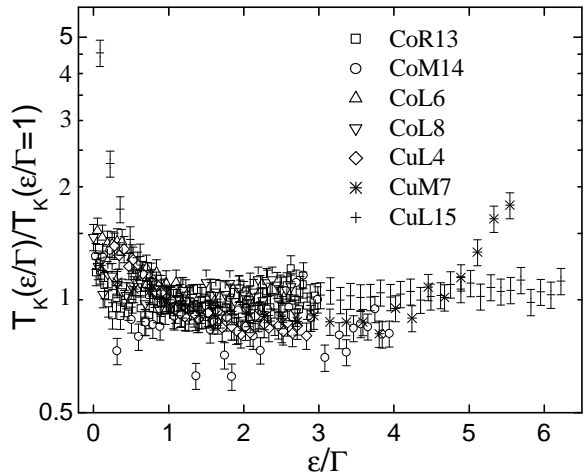}
\end{center}
\vspace{-5mm}
\caption{\small Log $T_{\rm K}$ (normalized) inferred from FWHM of
5~K Kondo peak in $G(V_{\rm SD})$, as a function of $V_{\rm G}$
normalized by the width of the Coulomb blockade charge degeneracy
point ($= \epsilon/\Gamma$ in the simple model of Fig.~\ref{fig1}a)
for several devices.  In the model of Fig.~\ref{fig1}a,
a parabolic dependence of log $T_{\rm K}$ on $V_{\rm G}$ is 
expected.  Values of $\Gamma$ inferred for these devices are,
top down, 22.6, 11.5, 18, 3.3, 12.6, 14.2, and 26.8 meV.
}
\label{fig3}
\vspace{-5mm}
\end{figure}

Figures~\ref{fig2}c,d are histograms of Kondo temperatures inferred
from resonance widths for 26 Co-containing Kondo devices (out of 921
electrode pairs examined at low temperatures), and 12 Cu-containing
Kondo devices (out of 397 electrode pairs examined at low
temperatures).  Only devices exhibiting zero-bias resonances and clear
charge degeneracy points are considered here.  The observed weak
dependence of $T_{\rm K}$ on $V_{\rm G}$ (see Fig.~\ref{fig3} and
later discussion) means that these distributions are relatively
insensitive to the choice of $V_{\rm G}$ at which $T_{\rm K}$ is
inferred.  No Kondo resonances were observed in 370 control devices
using alkanethiol chains, bare metal electrodes, and electrodes exposed to
solvents and poor vacuum.

Kondo physics in these complexes is clearly strong, with $T_{\rm K}$
values similar to those reported in STM measurements of Co atoms on
Au(111)\cite{MadhavanetAl98Science}.  The Kondo temperature is expected
to be\cite{vanderWieletAl00Science,Haldane78PRL}:
\begin{equation}
k_{\rm B}T_{\rm K} = \frac{\sqrt{\Gamma E_{\rm c}}}{2}e^{-\pi \epsilon(-\epsilon+E_{\rm c})/\Gamma E_{\rm c}}
\label{eq:TK}
\end{equation}
outside the mixed valence ($\epsilon/\Gamma < 1$) regime.  For small
molecules, $E_{\rm c}$ is likely to be hundreds of meV, while $\Gamma$
is empirically tens of meV.  The prefactor in (\ref{eq:TK}) then
implies $T_{\rm K}$ can be as large as hundreds of Kelvin.

Assuming a model as in Fig.~\ref{fig1}a, $\epsilon = 0$ at the charge
degeneracy point.  Normalizing the difference in $V_{\rm G}$ away from
charge degeneracy by the width in $V_{\rm G}$ of the charge degeneracy
resonance in the Coulomb blockade regime gives a lower limit on
$\epsilon/\Gamma$.  Figure~\ref{fig3} shows (normalized) $T_{\rm K}$
as inferred from low-$T$ resonance FWHM$~\sim~2k_{\rm B}T_{\rm K}/e$
as a function of inferred $\epsilon/\Gamma$ for several devices.
Eq.~(\ref{eq:TK}) predicts a quadratic dependence of $\log T_{\rm K}$
on $\epsilon$, with a minimum in $T_{\rm K}$ at $\epsilon = E_{\rm
c}/2$.  The measured dependence of inferred $T_{\rm K}(V_{\rm G})$ is
much less steep; indeed, for Sample CuM7, $T_{\rm K}$ actually {\em
increases} as $V_{\rm G}$ is shifted away from the nearest charge
degeneracy.  Thus, the simple model of Fig.~\ref{fig1}a, which works
well for semiconductor quantum dot
experiments\cite{GGetAl98PRL,vanderWieletAl00Science}, is inconsistent
with this data.  We note that the $T_{\rm K}(V_{\rm G})$ dependence
reported in Ref.~\cite{LiangetAl02Nature} for $\epsilon/\Gamma > 1$ is
also surprisingly weak.

We consider explanations for this deviation from simple expectations.
We dismiss as unlikely that the zero-bias peak may not be a true Kondo
resonance, given (a) quantitative consistency of the functional forms
for the resonance $G(V_{\rm SD},T)$ with Kondo expectations; (b) the
appearance of the zero-bias resonance coincident with passing through
charge degeneracy points; (c) the lack of such resonances in control
devices; (d) the similarity to other Kondo data reported in
SMTs\cite{ParketAl02Nature,LiangetAl02Nature}. 

The natural explanation for the anomalous gate dependence is that the
normalized $\delta V_{\rm G}$ used as $\epsilon/\Gamma$ in
Fig.~\ref{fig3} is not the true $\epsilon/\Gamma$ relevant to the
Kondo Hamiltonian that gives Eq.~(\ref{eq:TK}).  In semiconductor
quantum dots\cite{GGetAl98PRL} inferring $\epsilon/\Gamma$ by
normalizing $\delta V_{\rm G}$ is quantitatively consistent with
Eq.~(\ref{eq:TK}).  Presumably some mechanism intrinsic to the molecular
system renormalizes either the effective $\epsilon$, the effective
$\Gamma$, or both, away from the simple picture of Fig.~\ref{fig1}a.

\begin{figure}[h!]
\begin{center}
\includegraphics[clip, width=8cm]{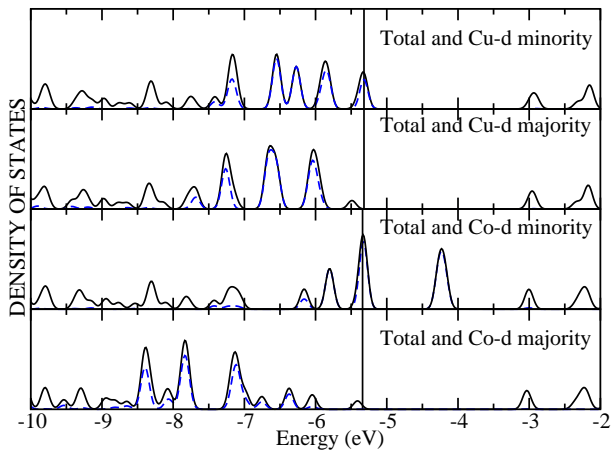}
\end{center}
\vspace{-5mm}
\caption{\small Spin-resolved projected densities of states of the
Cu(II) and Co(II) complexes, respectively.  Minority (upper) and
majority (lower) spin levels are indicated, demonstrating the strong
effects of intramolecular exchange.  The black curve is the total DOS,
and the dashed (blue) is the DOS projected onto the transition metal
$d$ states.  Further details are available\protect{\cite{supp}}.}
\label{fig4}
\vspace{-5mm}
\end{figure}

Orbital degeneracy for the unpaired spin is one
possibility\cite{BoncaetAl93PRB}.  When an $N$-fold degeneracy exists,
$T_{K}$ as defined by Eq.~(\ref{eq:TK}) is enhanced, such that the
denominator of the exponent becomes $N \Gamma E_{\rm c}$.  Thus the
normalization of the abcissa in Fig.~\ref{fig3} would effectively be
too large by a factor of $N$, qualitatively explaining the apparently
weak dependence of $T_{K}(V_{G})$ if $N \sim 5$.  However, EPR spectra
of both the Co and Cu-based complexes in solution phase show no
indication of such a large degeneracy.  Furthermore, we have performed
spin-resolved calculations on the transition metal complexes to
examine their electronic structure, as shown in Fig.~\ref{fig4}
(further details are available\cite{supp}).  In neither complex are
large degeneracies expected.  While it is conceivable that the
self-assembled compounds could have different orbital degeneracies
than the isolated molecules, it seems unlikely that both complexes,
with their differing isolated electronic structures, would have such
similar properties.

The electronic structure calculations reveal an additional energy
scale that is often small in semiconductor dots, but in the molecular
system is comparable energetically to the inferred $\Gamma$, $E_{\rm
c}$, $\epsilon$, and expected single-particle level spacing:
intramolecular exchange.  In both Co and Cu complexes, intramolecular
exchange is strong.  In the Co case, to a good approximation ligand
field effects split the $d$-states into one doubly degenerate, one
nearly doubly degenerate, and one singly degenerate set of states.
The complex has a total spin of $S=3/2$, as confirmed by EPR. A
minority spin half-occupied $d$ state is at the Fermi level with two
nearly degenerate $d$ states 1.1~eV above the Fermi level and the
remaining occupied minority $d$ electron 0.5~eV below the Fermi level.
Exchange splittings pull the majority $d$-band down significantly,
with the highest occupied majority $d$ state 1.8~eV below the Fermi
level. The fermiology is complicated by an additional delocalized
unpolarized two-fold degenerate carbon 2$p$ state lying 0.05~eV below
the Fermi level. Also there is a delocalized doubly degenerate
molecular state 0.85~eV below the Fermi level.  The Cu complex has a
total spin of $S=1/2$, with the Fermi level unchanged in comparison to
Co. This means that the minority spin $d$ states are pulled to lower
energies by approximately 1.1~eV as shown in the figure.  The
structure calculations also confirm significant delocalization of the
majority spin states, extending into the ligands, consistent with
large $\Gamma$ values for these systems.  Jahn-Teller distortion is
not considered, since the substrate-complex interaction is expected to
be large compared to this effect.

The occasional observation of Kondo resonances with similar $T_{\rm
K}$ values on {\it both} sides of charge degeneracy points further
indicates that intramolecular exchange cannot be neglected in these
devices.  However, it is not immediately clear how this strong
interaction could lead to the observed phenomenon.

Vibrational effects are another piece of physics intrinsic to the
molecular system.  The signature of electron-phonon interactions has
been observed in inelastic electron tunneling spectra of SMTs made
with these complexes\cite{YuetAl04PRL}.  Moderate coupling of charge
and vibrational modes\cite{CornagliaetAl04PRL} localized to the
molecule can strongly both increase $T_{\rm K}$ and decrease its gate
dependence relative to the case with no vibrational
coupling\cite{Grempel}.  A quantitative estimate of
electron-vibrational couplings would facilitate testing this
hypothesis, and should be obtainable from further quantum chemistry
calculations.

Finally, it is also possible\cite{SietAl93PRB} that screening
correlations in the mixed valence regime can renormalize the measured
$\Gamma$ to a value different than the $\Gamma$ relevant to the Kondo
temperature.  Full quantum chemistry calculations of molecules bound
to realistic Au leads including Kondo and many-body correlations (as
done for Co atoms on Au(111)\cite{UjsaghyetAl00PRL}) are essential to
a better understanding the observed effects, and are beyond the scope
of this paper.

In measuring the electronic properties of single-molecule transistors
containing transition metal complexes, we observe strong Kondo
physics, indicating that conjugated ligands can provide extremely
effective coupling of spin degrees of freedom to metal leads.  We also
find a $T_{\rm K}(V_{\rm G})$ that is vastly weaker than that seen in
semiconductor quantum dot realizations of the Kondo effect and
expected for the simple model of Fig.~\ref{fig1}a.  We have discussed
possible explanations for this anomalous dependence in light of
electronic structure calculations of the complexes.
While a complete understanding will require more sophisticated
modeling and further measurements, these data demonstrate that
correlated states involving SMTs can exhibit rich effects not seen in
their semiconductor quantum dot counterparts.

DN acknowledges support from the Research Corporation, the Robert
A. Welch Foundation, the David and Lucille Packard Foundation, an
Alfred P. Sloan Foundation Fellowship, and NSF award DMR-0347253.  JMT
acknowledges support from DARPA and the ONR.  MRP and TB acknowledge
ONR (Grant No. N000140211046) and DoD CHSSI program.  TB acknowledges
NSF award NIRT-0304122.  The authors thank Dr. M. Fabian and
Prof. G. Palmer for EPR characterization, Dr. M.P. Stewart for XPS
characterization, and Prof. Q. Si for a critical reading of the
manuscript.

\end{document}